\documentclass[12pt,preprint]{aastex}

\lefthead{K. Krisciunas {\em et al.}}
\righthead{Hubble Diagrams of Type Ia Supernovae}

\begin{document} 

\title{Hubble Diagrams of Type Ia Supernovae in the Near Infrared}

\author{Kevin Krisciunas,$^{1,2}$
Mark M. Phillips,$^1$ and
Nicholas B. Suntzeff$^2$}
\affil{$^1$Las Campanas Observatory, Carnegie Observatories, Casilla 601, La
Serena, Chile \\
$^2$Cerro Tololo Inter-American Observatory, National Optical
Astronomy \\ Observatory,\altaffilmark{3} Casilla 603, La Serena, Chile} 
\altaffiltext{3}{The National Optical Astronomy
Observatory is operated by the Association of Universities for
Research in Astronomy, Inc., under cooperative agreement with the
National Science Foundation.}

\email {kkrisciunas, nsuntzeff@noao.edu \\
mmp@lco.cl }

\begin{abstract}

From observations of 7 Type Ia supernovae obtained during the last
four years at the Las Campanas and Cerro Tololo Inter-American
Observatories, along with previous published data for 9 supernovae, we
present $JHK$ Hubble diagrams and derive absolute magnitudes at
maximum light of 16 objects out to a redshift of 0.038.  On the scale
of H$_0$ = 72 km s$^{-1}$ Mpc$^{-1}$ we find mean absolute magnitudes at maximum
of $-$18.57, $-$18.24, and $-$18.42 for $J$, $H$, and $K$,
respectively, with 1-$\sigma$ uncertainties of the distributions of
values of $\pm$ 0.14, 0.18, and 0.12 mag.  The data indicate no
significant decline rate relations for the infrared.  Thus, Type Ia
SNe at maximum brightness appear to be {\em standard candles} in the
infrared at the $\pm$ 0.20 mag level or better.  The minimum requirements
for obtaining the distance to a Type Ia SN are: reasonably
accurate values of $\Delta$m$_{15}$($B$) and T($B_{max}$), and one night
of infrared data in the $-$12 to +10 d window with respect to T($B_{max}$).
\end{abstract}

\parindent = 0 mm

\keywords{supernovae, photometry; supernovae}

\section{Introduction}

\parindent = 9 mm

Type Ia supernovae (SNe) are the most precise distance indicators for
extragalactic astronomy at redshift $z \ga 0.01$.  Since the
discovery that the absolute magnitudes at maximum light were related
to the decline rates (Pskovskii 1977, 1984; Phillips 1993), Type Ia
SNe have been considered standardizable candles.  The intrinsically
brighter ones have wider $B$- and $V$-band light curves, along with
stronger and later secondary humps in the $I$-band light curves.  The
three principal schemes for characterizing the light curves are the
$\Delta$m$_{15}$ method (Hamuy et al. 1996; Phillips et al.  1999),
the multicolor light curve shape (MLCS) method (Riess, Press, \&
Kirshner 1996; Riess et al. 1998), and the ``stretch method'' of
Perlmutter et al.  (1997).

Elias et al. (1981, 1985) were the first to publish any extensive
infrared (IR) photometry of Type Ia SNe.  In their Fig. 6, they
presented the first $H$-band Hubble diagram of Type Ia SNe and
commented, ``the dispersion in their absolute magnitude near maximum
is small, making them potentially useful distance indicators.''  After
this, with the exception of SN~1986G (Frogel et al. 1987), very few IR
observations were made of Type Ia SNe until the appearance of
SN~1998bu (Jha et al. 1999, Hernandez et al. 2000). 

Meikle (2000) gives a summary of the IR data available 4 years ago and states
that ``most SNe Ia are {\em standard} IR candles at a level of about 0.15 mag.''
Meikle's analysis related to the absolute magnitudes in $JHK$ at 14 days after
T($B_{max}$) and used host galaxy distance determinations via Cepheids observed
with HST. In this paper we use four of the same objects considered by Meikle,
namely SNe~1980N, 1981B, 1986G, and 1998bu.  We key on their apparent and
absolute magnitudes at {\em maximum} light.  The majority of our objects are
distant enough to be in the smooth Hubble flow.

It is well known that light is less extinguished by dust at infrared
wavelengths compared to optical wavelengths.  According to Cardelli,
Clayton, \& Mathis (1989), A$_{\lambda}$/A$_V$ = 0.282, 0.190, and
0.114, for the near-IR $JHK$-bands.  In principle, the near infrared
bands should be less subject to systematic errors due to dust
extinction along the line of sight.  Until recently it has been
difficult to test this hypothesis because few Type Ia SNe have been
observed early enough to overlap the times of infrared maximum, which
generally occur about 3 days prior to the $B$-band maximum.

Simple experimentation on the data showed us that the rest frame light
curves could be fit by template $JHK$ light curves (Krisciunas et al.
2004a, \S 3.1) using the stretch technique of Perlmutter et al. (1997)
provided that the fits were made in the window of time $-$12 to +10 d with
respect to T($B_{max}$).  To create the near-IR templates, the data were
transformed to the SN rest frame by applying K-corrections to the observed
magnitudes and time dilation corrections to the time from maximum light.  
Since we did not have the Perlmutter templates or the code to calculate
the stretch factors directly, we estimated the optical stretch factors
using the relationships of Jha (2002, Fig. 3.8), which relate the $B$- and
$V$-band stretch factors to the Phillips parameter $\Delta$m$_{15}$($B$).  
We have very few IR light curves that are well sampled and cover the maxima.
To bring all the data to a standard $s=1$ stretch, we used the inverse
stretch factors $s^{-1}$ (averaging the $B$- and $V$-band values) to scale
the time axis of the IR light curves. The final templates, shown in Fig.
1, represent the averaged behavior in $JHK$ for a supernova with
$s=1$.\footnote[4] {We shall define ``stretched time'' ($t^{\prime}$) to
be the number of days with respect to T($B_{max}$), multiplied by the
inverse stretch factor $s^{-1}$ and divided by (1 + $z$).} Our $JHK$
templates exhibit {\sc rms} deviations of $\sigma_J = \pm$ 0.062,
$\sigma_H = \pm$ 0.080, and $\sigma_K = \pm$ 0.075 mag around a cubic fit.  
Except at $t^{\prime} \; \gtrsim$ +7 d in the $H$-band diagram, these
uncertainties seem to be {\em bona fide} measures of the {\sc rms}
errors, rather than evidence for systematic differences between objects.
For $RIJHK$ the stretch method does not work beyond $t \; \approx$ +10 d because
the secondary humps do not scale like the maxima.

We then used the adopted maximum apparent magnitudes of eight of the
nine template SNe, estimates of the $V$-band extinction appropriate to
each object, and the ratios of IR extinction to A$_V$ of Cardelli et
al. (1989) to obtain extinction-corrected maxima.  For the seven objects
with sparser data, we used the templates and appropriate
extinction corrections to determine their extinction-corrected maxima.
Finally, we include the unusual SN~1999ac (Phillips et al. 2003),
which was reasonably well sampled at the time of its IR maxima.

In this letter we present the $JHK$ Hubble diagrams and absolute
magnitudes at maximum light  based on the data in Krisciunas et
al. (2004a).

\section{The Data}

We consider the following Type Ia SNe:

\parindent = 0 mm

SN~1980N (Elias et al. 1981).  For this SN and its host, NGC 1316,
we use the weighted mean of the two distance moduli given by Ajhar
et al. (2001, Table 3), $m-M = 31.44 \pm 0.14$.  This distance
modulus is based on surface brightness fluctuations (SBFs) of the
host and is on the Cepheid scale of Freedman et al. (2001), with
H$_0$ = 72 km s$^{-1}$ Mpc$^{-1}$. 



SN~1981B (Elias et al. 1981, 1985; Rafanelli et al. 1981;
Salinari \& Moorwood 1981).  Freedman et al. (2001) give a
Cepheid distance modulus of $\mu_0$ = 30.80 $\pm$ 0.04 for the host
(NGC 4536).

SN~1994D was observed by Richmond et al. (1995) in the IR on only
one night near maximum.  Ajhar et al. (2001, Table 3) give an SBF
distance modulus of $m-M = 31.08 \pm 0.20$.  

SN~1986G (Frogel et al. 1987).  Phillips et al. (1999) indicate that
E($B-V$)$_{host}$ = 0.50 mag. The Schlegel, Finkbeiner, \& Davis
(1998) Galactic reddening is E($B-V$)$_{Gal}$ = 0.115. From
polarimetry Hough et al. (1987) indicate that R$_V \equiv$ A$_V$ /
E($B-V$) = 2.4 $\pm$ 0.13 for the dust in the host galaxy, NGC 5128,
much less than the nominal Galactic value of R$_V$ = 3.1.  For this
object A$_V$ = 3.1 $\times$ 0.115 + 2.4 $\times$ 0.50 = 1.56, to
which we assign a large uncertainty. Ajhar et al. (2001, Table 3)
give a distance modulus of $m-M = 28.06 \pm 0.14$ on the basis of
SBFs.  

SN~1998bu (Suntzeff et al. 1999; Jha et al. 1999; Hernandez et al.
2000). We exclude the data of Mayya, Puerari, \& Kuhn (1999), which we
consider uncertain.  Freedman et al. (2001, Table 4) give a distance
modulus of $\mu_0$ = 29.97 $\pm$ 0.06 for the host, NGC 3368, on the
basis of their final Cepheid calibrations.
  
SN~1999aa (Krisciunas et al. 2000) and SN~1999gp (Krisciunas et al.
2001).  We assume that they are unreddened in their hosts.  We use a
corrected decline rate of $\Delta$m$_{15}$($B$) = 0.81 $\pm$ 0.04 for
SN~1999aa (J. L. Prieto, private communication), which is based on the
photometry of Krisciunas et al. (2000) and Jha (2002).  SN~1999aa at
optical wavelengths was similar in many ways to SN~1999aw.

SN~1999ac (Phillips et al. 2003).  This object was spectroscopically
peculiar and its $B-V$ colors were unusual in the tail of the color
curve.  Because of the peculiarities of SN~1999ac, we have not used it
as a template object, but the $K$-band data in particular are very
similar to our $K$-band template.

SN~1999aw (Strolger et al. 2002).  The $H$- and $K$-band photometry
has lower S/N due to the higher redshift (z=0.038) of the object.  We
did not use the $K$-band data of this object for the construction of
our $K$-band template. The host galaxy is of very low surface
brightness and we assume no host reddening.  This
object was similar to SN~1999aa. Both were slow decliners and
exhibited spectroscopic peculiarities.

SN~1999cp (Krisciunas et al. 2000).  MLCS fits indicate that his object
was unreddened in its host.  $\Delta$m$_{15}$ analysis gives
E($B-V$)$_{host}$ = 0.04 $\pm$ 0.03.  We adopt the MLCS value
of T($B_{max}$) because the MLCS fitting is able to include the
earliest photometric points in the fit and is therefore able to
constrain T($B_{max}$) more precisely.

SN~1999ee.  Stritzinger et al. (2002) give A$_V$ = 0.94 $\pm$ 0.16 from
optical data.  Krisciunas et al. (2004a) give extensive IR data.

SN~2000bh.  Krisciunas et al. (2004a) give IR data on 23 nights
starting at $t^{\prime}$ = +5.5 d.  The sparser optical data begin
one day later.  

SN~2000bk (Krisciunas et al. 2001).  Minimally reddened in its host.

SN~2000ce. This object is considerably reddened in its host. MLCS gives
A$_V = 1.67\; \pm$ 0.20 mag (Krisciunas et al. 2001, Table 14 and \S3.2).

SN~2000ca and SN~2001ba.  See Krisciunas et al. (2004a).  Neither is
highly reddened in its host.

SN~2001el (Krisciunas et al. 2003). Though this object had normal
light curves and was well sampled, it is not far enough to be in the
quiet Hubble flow.  Lacking a directly measured distance to the host
galaxy via Cepheids or SBFs, we reluctantly eliminate it from further
consideration of this paper.

\parindent = 9 mm

The uncertainties of the magnitudes at maximum light are given by the
photometry alone for the objects with well sampled light curves at maximum.  
For the other objects the uncertainties of the apparent magnitudes at maximum
were taken to be the square root of the quadratic sum of the typical
uncertainties of the photometry and the uncertainties of our template
{\sc rms} values ($\sigma_J$, $\sigma_H$, $\sigma_K$).

In order to make Hubble diagrams we need corresponding velocities in
the frame of the Cosmic Microwave Background (CMB).  Ideally, we would
only consider objects in the Hubble flow ($v \gtrsim 3000$ km
s$^{-1}$) to reduce the effects of peculiar
velocities on the dispersion in the flow.  Five of our objects are
closer than this.

For two of the nearest objects (SNe~1981B and 1998bu) we used the Cepheid
distances of Freedman et al. (2001). For SNe~1980N, 1986G, and 1994D we
used SBF distances on the Cepheid scale of Freedman et al. (2001) and
simply multiplied the distances by H$_0$ = 72 km s$^{-1}$ Mpc$^{-1}$ to
get ``equivalent'' velocities in the CMB frame.\footnote[5]{We also
employed the flow model of Tonry et al. (2000) to calculate the equivalent
distances of the five nearest objects in our sample which give flow
velocities equal to the CMB velocities from NED.  This gave equivalent
distances of SNe~1986G, 1980N, and 1998bu in reasonble agreement with the
directly determined values. However, the model is too simplistic in the
cases of SNe~1981B and 1994D, which are found in direction of the Virgo
cluster itself.} For the other 11 objects we used the redshifts in the CMB
frame.

In Table 1 we give the values of $\Delta$m$_{15}$($B$), the velocities
in the CMB frame or their equivalents, values of A$_V$ derived from
optical light curves, the adopted $JHK$ maxima of the SNe and their
maxima corrected for extinction along the line of sight.  For those
objects assumed unreddened in their hosts we adopted the values of
Galactic reddening E($B-V$) of Schlegel et al. (1998) and assumed
R$_V$ = 3.1.  To correct the observed IR maxima for dust extinction we
assumed the A$_{\lambda}$/A$_V$ ratios of Cardelli et al. (1989).

\section{Results}
   
In Fig. \ref{jhk_hubble} we show the Hubble diagrams of Type Ia
SNe for the near-IR $JHK$ bands, plotting the
extinction-corrected apparent magnitudes versus the logarithm of the
red\-shifts in the CMB frame.  For the five objects with directly
measured distances we plot the ``equivalent'' redshifts.  

It is obvious from an inspection of Fig. \ref{jhk_hubble} that our
sample of Type Ia SNe may be regarded as excellent standard candles
in the IR.  SNe~1999ee and 1999cp, one well sampled and reddened, the
other sparsely sampled and unreddened, with nearly identical
velocities, fall on top of each other in all three sub-diagrams.  We
note that while SNe~1999aa, 1999ac, and 1999aw are ``peculiar'' in
some ways at optical wavelengths, there seems to be nothing peculiar
about their IR luminosities compared to other objects. 

The next question to ask is clearly: are the deviations from the
Hubble lines in Fig. \ref{jhk_hubble} a function of
$\Delta$m$_{15}$($B$)?  In other words: are there decline rate
relations for the IR bands?  To answer this we derive the absolute
magnitudes of the objects in our sample.

All of our absolute magnitudes were determined on the Freedman et al.
(2001) scale of H$_0$ = 72 km s$^{-1}$ Mpc$^{-1}$. For distances, we
used either the directly measured distances for the hosts of
SNe~1980N, 1981B, 1986G, 1994D, and 1998bu or CMB-frame velocities
converted to distances via Hubble's Law.  In Fig. \ref{absmags} we
show the $JHK$ absolute magnitudes of our sample.  The uncertainties
in the absolute magnitudes include the random errors of the
photometry, template fitting, extinction corrections, and also the
distance moduli.  For those objects in the Hubble flow we assumed a
representative peculiar motion of $\pm$ 300 km s$^{-1}$.  This
translates to an uncertainty in the distance modulus of 0.222 mag for
SN~1999ac, but only 0.055 mag for SN~1999aw.  From Fig. \ref{absmags}
we deduce that there are no obvious decline rate relations in the
near-IR.

In Table 2 we give the weighted mean values and dispersions of the
absolute magnitudes at maximum.  Column 3 of this table contains the
1-$\sigma$ Gaussian widths of the distributions of the absolute
magnitudes.  In Table 2 we also give the reduced $\chi^2$ values,
which (since they are close to 1) demonstrate that we have assumed
sensible uncertainties for the adopted distance moduli and photometry,
and that the absolute magnitudes are constant within the errors.

Figs. \ref{jhk_hubble} and \ref{absmags} also demonstrate the
soundness of the assumption that a single Hubble constant 
is applicable out to a redshift of 0.04.  Otherwise there would
be a shift with respect to the lines of slope 5 in Fig.
\ref{jhk_hubble}, and the points corresponding to nearby objects
and those in the Hubble flow in Fig. \ref{absmags} would exhibit
some kind of systematic difference.

While Type Ia SNe are {\em standardizable} candles in the optical
bands, they apparently are {\em standard} candles in the near-IR, at
the $\pm$ 0.20 mag level or better ($\pm$ 9\% in distance), depending
on the filter.  This confirms the prognostication of Elias et al.
(1985) and the analysis of Meikle (2000), which had the benefit
of Cepheid-based distance determinations of the host galaxies of
8 objects.\footnote[6]{In this paper we have considered 16 objects,
11 of which are distant enough to be in the smooth Hubble flow.
At the time of this update, 1 March 2004, we have 5 more
objects we can add to the diagrams: SNe 1999ek, 2001bt, 2001cn,
2001cz, and 2002bo.  The first 4 of these are in the smooth Hubble
flow.  See Krisciunas et al. (2004b) for further details. The 
reader could also consider the more problematic cases of SNe~2000cx 
(Candia et al. 2003) and 2001el (Krisciunas et al. 2003).}

In this paper we have outlined a new method of determining distances
to Type Ia SNe.  Of course, the use of highly non-standard $JHK$ filters
is to be avoided.  Our method requires a minimum of one night's IR data
in the window $-12$ to +10 d with respect to T($B_{max}$), and reasonably
accurate values of T($B_{max}$) and $\Delta$m$_{15}$($B$).  This exploits
the nature of the IR light curves, which are well behaved and obey the
stretch model {\em at maximum light}.  By focusing on IR light curves we
employ extinction corrections whose values and uncertainties are much
smaller than in the optical.  Not only does this lead to very small
scatter in the near-IR Hubble diagrams, but it underscores the standard
candle nature of Type Ia SNe in the IR.  


\acknowledgements  

We gratefully made use of the NASA/IPAC Extragalactic Database
(NED). This publication makes use of data products from the 2MASS
Survey, funded by NASA and the NSF.
KK and NBS thank STScI for their support through grants
HST-GO-07505.02-96A, HST-GO-08641.07-A, and HST-GO-09118.08-A.  We thank
Mario Hamuy, Darren DePoy, and Peter Meikle for scientific comments on
this work. We thank Pablo Candia, Sergio Gonzalez, and Jose Luis Prieto
for help on the data reduction. We thank Eric Persson and Darren DePoy for
providing the excellent infrared instrumentation used in this study. Some
of the data used in this study were obtained at CTIO using the``Small and
Moderate Aperture Research Telescope System'' (SMARTS).  We thank John
Tonry for a copy of his flow model program and for useful discussions. KK
thanks LCO and NOAO for funding part of his postdoctoral position.

\clearpage


\begin{deluxetable}{lclcccccccc}
\tablecolumns{11}
\tabletypesize{\scriptsize}
\rotate
\tablewidth{0pc}
\tablecaption{Data for Type Ia Supernovae}
\startdata
SN & $\Delta$m$_{15}$($B$) & $v_{CMB}$ & A$_V$ & $J_{max}$ & $J_{corr}$ &
$H_{max}$ & $H_{corr}$ & $K_{max}$ & $K_{corr}$ & N$_{J,H,K}^a$\\ 
 & & (km s$^{-1}$) & & & & & & & &\\ \hline \hline
1980N$^b$  & 1.28(04) &   1397$^c$ &  0.22(06) & 12.84(08) & 12.78(08) & 13.24(10) & 13.20(10) & 13.10(10) & 13.08(10) & 3,3,3 \\  
1981B      & 1.10(07) &   1041$^c$ &  0.40(09) & 12.44(08) & 12.33(08) & 12.88(10) & 12.81(10) & 12.61(09) & 12.57(09) & 3,3,3 \\
1986G$^b$  & 1.73(07) &   295$^c$ &  1.56(40) & 10.00(03) & \phn9.56(12) & \phn9.96(03) & \phn9.66(10) & \phn9.85(03) & \phn9.67(06) & 5,6,6 \\
1994D      & 1.31(08) &   1184$^c$ &  0.12(02) & 12.65(09) & 12.62(09) & 12.65(10) & 12.63(10) & 12.59(10) & 12.58(10) & 1,1,1 \\
1998bu$^b$ & 1.01(05) &   710$^c$ &  1.13(20) & 11.65(06) & 11.33(08) & 11.77(10) & 11.56(11) & 11.55(05) & 11.42(06) & 11,11,11 \\
1999aa     & 0.81(04) &   4572 &  0.12(01) & 15.65(07) & 15.62(07) &  \nodata  &   \nodata   & 15.66(10) & 15.65(10) & 1,0,1  \\
1999ac     & 1.34(08) &   2943 &  0.51(20) & 14.53(03) & 14.39(06) & 14.68(03) & 14.58(05) & 14.56(03) & 14.50(04) & 14,14,14\\
1999aw$^b$ & 0.81(03) &  11750 &  0.10(02) & 17.71(06) & 17.68(06) & 18.00(16) & 17.98(16)  & 17.65(11) & 17.64(11)  & 9,8,9 \\
1999cp     & 0.87(10) &   3115 &  0.07(01) & 14.55(02) & 14.53(02) & 14.78(02) & 14.77(02) & 14.61(06) & 14.60(06) & 2,2,2 \\
1999ee$^b$ & 0.94(06) &   3163 &  0.94(16) & 14.78(04) & 14.52(06) & 15.02(04) & 14.84(05) & 14.62(03) & 14.51(04) & 17,18,6 \\
1999gp     & 1.00(10) &   7806 &  0.17(03) & 16.41(09) & 16.36(09) & 17.16(15) & 17.13(15) & 16.75(18)  & 16.73(18) & 1,1,2 \\
2000bh     & 1.16(10) &   7181 &  0.21(09) & 16.39(08) & 16.33(08) & 16.75(09) & 16.71(09) & 16.66(09)  & 16.64(09) & 5,5,5 \\
2000bk     & 1.63(10) &   7976 &  0.28(20) & 16.87(07) & 16.79(09) & 17.39(09) & 17.34(10) &   \nodata   &   \nodata   & 2,2,0 \\
2000ca$^b$ & 0.98(05) &   7352 &  0.21(09) & 16.41(03) & 16.35(04) & 16.71(09) & 16.67(09) &   \nodata   &   \nodata   & 7,7,0 \\
2000ce     & 0.99(10) &   4946 &  1.67(20) & 15.85(07) & 15.38(09) & 16.27(08) & 15.95(09) & 15.97(09)  & 15.78(09) & 1,1,1 \\
2001ba$^b$ & 0.97(05) &   9152 &  0.32(09) & 17.04(03) & 16.95(04) & 17.26(03) & 17.20(03) & 17.22(06)  & 17.18(07) & 5,5,3 \\
\enddata
\tablenotetext{a} {Number of data points prior to $t^{\prime}$ = +10 d in ``stretched time''.
$^b$Objects used for $JHK$ templates.  SN~2001el was also a template object.  See text for further comments.
$^c$See text for comments on ``equivalent'' velocities.}
\end{deluxetable}

\clearpage

\begin{deluxetable}{ccccc}
\tablecolumns{4}
\tablewidth{0pc}
\tablecaption{Mean Absolute Magnitudes of Type Ia SNe at Maximum}
\startdata
Filter     &  $\langle$M$\rangle$ & $\sigma_x$ & $\chi^2_\nu$ & N     \\ \hline \hline
  J        &     $-$18.57(03)     &   $\pm$ 0.138  & 1.27 & 16  \\
  H        &     $-$18.24(04)     &   $\pm$ 0.183  & 1.67 & 15  \\
  K        &     $-$18.42(04)     &   $\pm$ 0.121  & 0.77 & 14  \\
\enddata
\end{deluxetable}

\clearpage


\figcaption[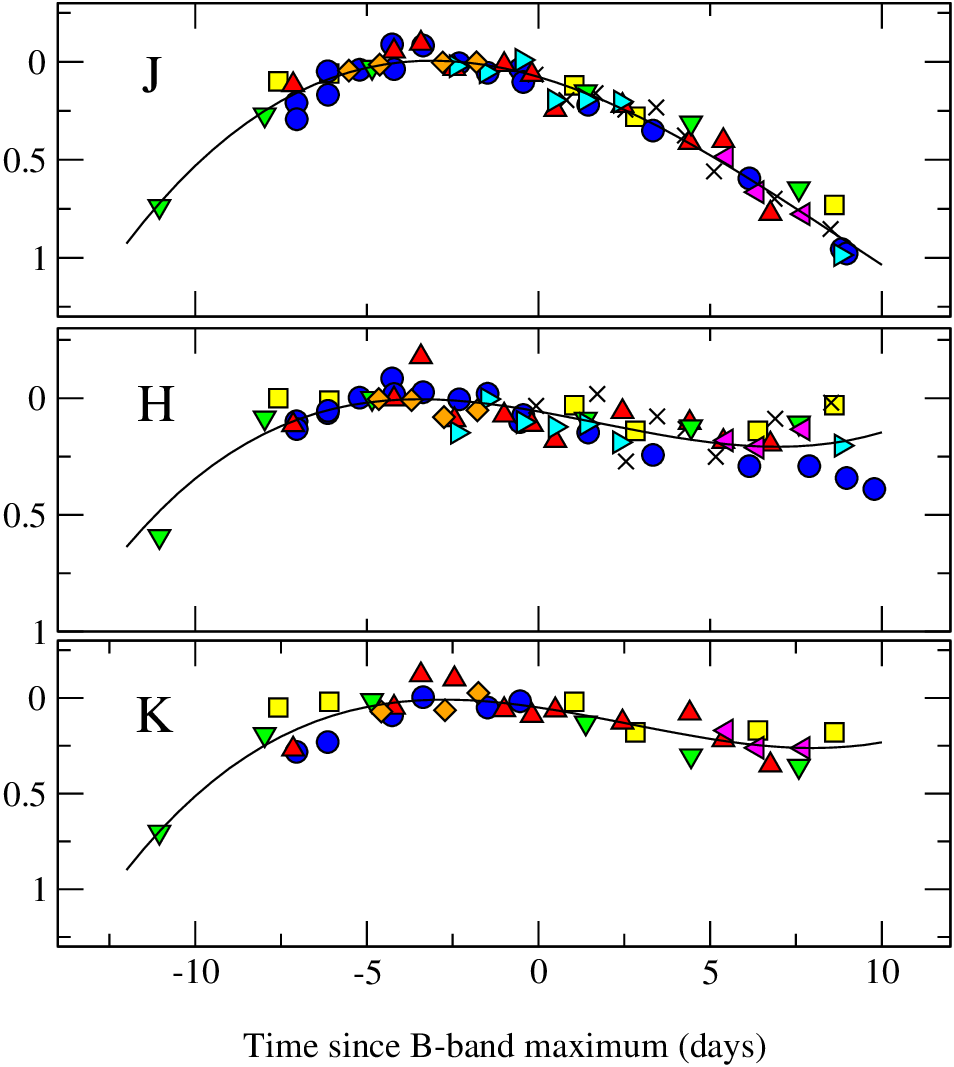] {$JH$ templates based on 8 Type Ia SNe and
$K$-band template based on 6 objects. The time axis is ``stretched
time'' (in days), which has been corrected for time dilation and scaled to a
fiducial $s=1$ stretch factor.  Third order fits to the data are shown.
The symbols for various supernovae are: 1980N (magenta triangles, pointing left), 
1986G (yellow squares), 1998bu (red triangles, pointing up), 1999aw (X's),
1999ee (blue circles), 2000ca (cyan triangles, pointing right,
2001el (green triangles, pointing down), 2001ba (orange diamonds).
\label{jhk_maxima} }

\figcaption[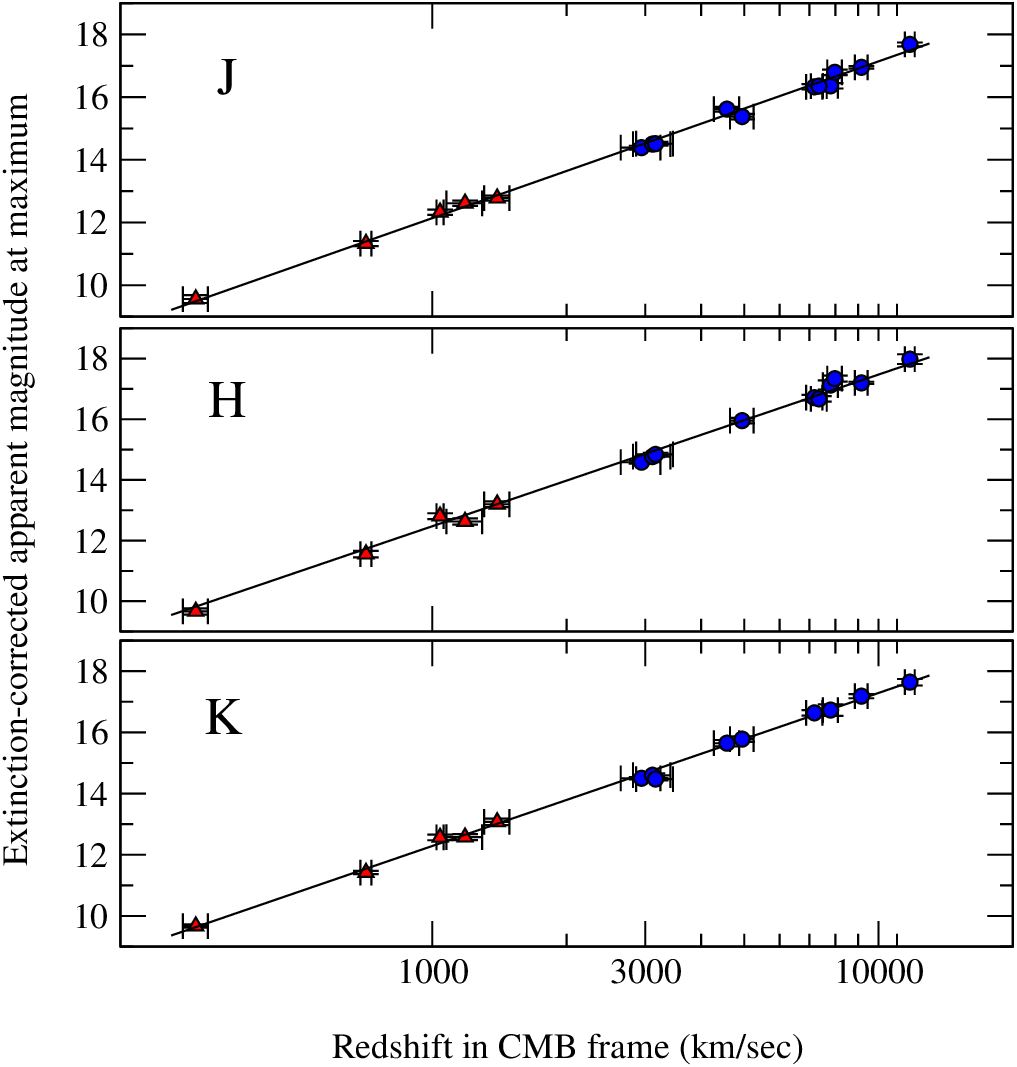] {Hubble diagrams of Type Ia SNe.  We plot
the extinction-corrected $J$-, $H$-, and $K$-band maxima versus the
logarithm of the redshifts in the CMB frame.  The points represented by 
blue circles have velocities taken from NED.  We have added horizontal error bars
corresponding to typical peculiar velocity of $\pm$ 300 km s$^{-1}$.
The points represented by red triangles have ``equivalent'' CMB
velocities derived from direct measures of the distances to the hosts,
on the H$_0$ = 72 km s$^{-1}$ Mpc$^{-1}$ scale of Freedman et
al. (2001). The straight lines given in each plot have a slope of
exactly 5 and correspond to the mean absolute magnitudes in each
filter on an H$_0$ = 72 scale. \label{jhk_hubble} }

\figcaption[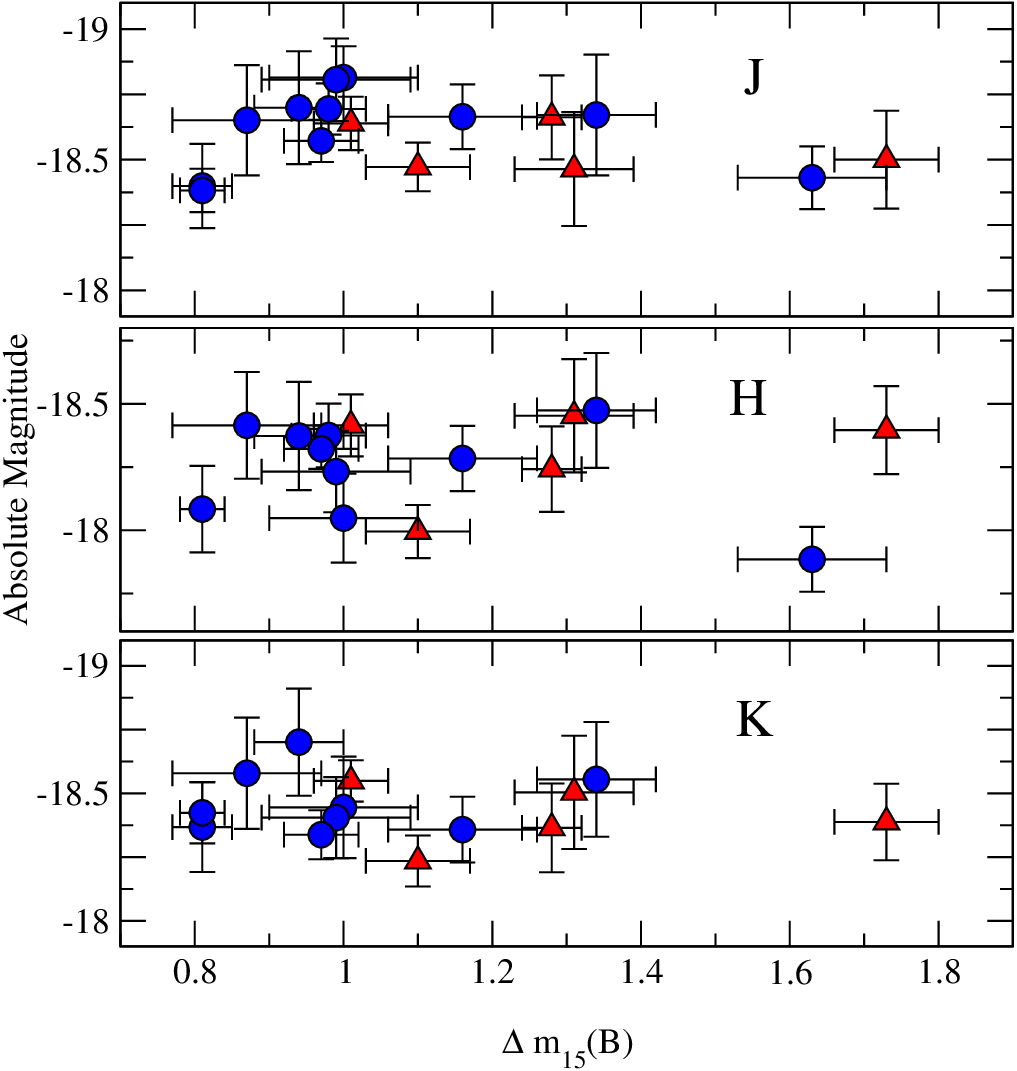] {Absolute magnitudes of Type Ia SNe at
maximum on the H$_0$ = 72 km s$^{-1}$ Mpc$^{-1}$ scale. Symbols:
blue circles are objects in the Hubble flow ($v \gtrsim$ 3000 km
s$^{-1}$); red triangles are SNe~1980N, 1981B, 1986G, 1994D, 1998bu
which have Cepheid or SBF distances.  The uncertainties in the
absolute magnitudes take into account the random errors of the
photometry, template fitting, extinction corrections, and the
determination of the distance moduli. \label{absmags} }

\clearpage

\begin{figure}
\plotone{jhk_maxima.eps}
{\center Krisciunas {\it et al.} Fig. \ref{jhk_maxima}}
\end{figure}

\begin{figure}
\plotone{jhk_hubble.eps}
{\center Krisciunas {\it et al.} Fig. \ref{jhk_hubble}}
\end{figure}

\begin{figure}
\plotone{absmags.eps}
{\center Krisciunas {\it et al.} Fig. \ref{absmags}}
\end{figure}


\end{document}